\documentstyle[aps,epsf,twocolumn,prl]{revtex}

\begin{document}
\def \tr{{\mbox{tr~}}}
\def \ra{{\rightarrow}}
\def \be{\begin{equation}}
\def \ee{\end{equation}}
\def \bea{\begin{eqnarray}}
\def \eea{\end{eqnarray}}
\def \nn{\nonumber}
\def \half{{1\over 2}}
\def \etal{{\it {et al}}}
\def \cH{{\cal{H}}}
\def \cM{{\cal{M}}}
\def \cN{{\cal{N}}}
\def \cQ{{\cal Q}}
\def \bS{{\bf S}}
\def \bL{{\bf L}}
\def \tJ{{\tilde{J}}}
\def \W{{\Omega}}
\def \e{{\epsilon}}
\def \lam{{\lambda}}
\def \a{{\alpha}}
\def \t{{\theta}}
\def \b{{\beta}}
\def \g{{\gamma}}
\def \D{{\Delta}}
\def \d{{\delta}}
\def \w{{\omega}}
\def \s{{\sigma}}
\def \f{{\varphi}}
\def \x{{\chi}}
\def \h{{\eta}}
\def \hatt{{\hat{\t}}}
\def \hn{{\bar{n}}}
\def \vk{{\bf{k}}}
\def \vq{{\bf{q}}}
\def \gk{{\g_{\vk}}}
\def \nd{{^{\vphantom{\dagger}}}}
\def \yd{^\dagger}
\def \ket#1{{\,|\,#1\,\rangle\,}}
\def \bra#1{{\,\langle\,#1\,|\,}}
\def \braket#1#2{{\,\langle\,#1\,|\,#2\,\rangle\,}}
\def \expect#1#2#3{{\,\langle\,#1\,|\,#2\,|\,#3\,\rangle\,}}
\def \rl#1#2{{\,\langle\,#1\,#2\,\rangle\,}}
\def \ad{{\a\yd}}
\def \an{{\a\nd}}
\def \av#1{{\langle#1\rangle}}
\def \bd#1{{(\sin\t\ad_{0#1}+\cos\t\ad_{1#1})}}
\def \bn#1{{(\sin\t\an_{0#1}+\cos\t\an_{1#1})}}
\def \sd#1{{(\cos\t\ad_{0#1}-\sin\t\ad_{1#1})}}
\def \sn#1{{(\cos\t\an_{0#1}-\sin\t\an_{1#1})}}
\draft
\twocolumn[\hsize\textwidth\columnwidth\hsize\csname @twocolumnfalse\endcsname

\title{Oscillating Superfluidity of Bosons in Optical Lattices
}
\author{Ehud Altman  and Assa Auerbach}
\address{Department of
Physics, Technion, Haifa 32000, Israel.} \date{\today} \maketitle
\begin{abstract}
Following a  suggestion by Orzel \etal\cite{orzel}, we analyze bosons in an optical lattice
undergoing  a sudden parameter change from the Mott to superfluid phase.
We introduce a modified coherent states path integral to
describe both phases.
The saddle point theory yields  collective oscillations of  the uniform
superfluid order parameter.
We calculate  its damping rate by phason pair emission. In two dimensions
the overdamped region largely overlaps with the quantum critical region.
Measurements of   critical dynamics  on  the Mott side are proposed.
\end{abstract} 

\vskip2pc]
\narrowtext
With recent  experimental developments of
ultra cold atoms in optical lattices, the fascinating phenomena of
Bose-Einstein Condensation  have  entered the domain of strong
interactions\cite{orzel,bloch}. Macroscopic quantum states can be effectively
manipulated and time evolution of order parameters (OP), 
adiabatic\cite{bloch} or
non-adiabatic\cite{orzel}, can be probed by varying the optical lattice
parameters.

In one such experiment, the strength
of a three dimensional optical lattice potential was tuned to
induce a quantum phase transition
between a Mott insulator and a superfluid of bosons\cite{bloch}.
This phase transition has been extensively analyzed theoretically
\cite{fisher,kotliar,caffarel,ramakrishnan,moonien}
and numerically\cite{QMC}.
The two phases are characterized by markedly different many body states.
The Mott phase, at large lattice potential barriers,
is well described by definite real space occupation numbers. 
The compressible superfluid phase,   on
the other hand, sustains long range  phase order. This  phase
is detected by self  interference patterns after the gas  is released from
the trap.

In an interesting proposal, Orzel \etal \cite{orzel} suggested  the
possibility of observing OP time evolution. Basically,
the bosons are prepared in  the  number
squeezed Mott state, and then   the   potential is suddenly reduced   into
the superfluid phase. The consequent evolution of the superfluid order can be
deduced from the intensity of interference
patterns appearing when the atoms are released from the trap
at sequential times.
This would open up exploration of a new regime of
{\em macroscopic quantum dynamics}\cite{kramer,franzosi,zurek}.
The initial  questions which come to mind
are: (i)  Could   coherence oscillations  be observed? (ii) What would be the
time scale of superfluid OP evolution?   (iii) What would be the
damping mechanism, and damping rate  of such  effects?

These are the primary issues addressed in this paper. We derive
the effective Hamiltonian of  the superfluid OP
starting from interacting bosons in  a periodic
lattice.  We find a variational bosonic  representation which describes the
phase diagram and treats the elementary excitations on both
the Mott and superfluid phases.  In the Mott phase, the  two degenerate gapped
excitations become gapless at the transition. The superfluid phase
reduces to a relativistic Gross-Pitaevskii  action, with  one  (gapless)
phase  mode and one (gapped)  amplitude   mode. We obtain a
semiclassical solution of a macroscopically oscillating superfluid order and
calculate its relative damping rate. This provides an
estimate of the experimental  regime  where such oscillations
should be visible. In two dimensions this region largely overlaps with the
quantum critical region, as we  estimate  from Ginzburg's
criterion. We end by commenting on critical dynamics, and how they might be
observed.

The Bose Hubbard model (BHM) describes
interacting  bosons in an optical lattice,
\be
H=\frac{U}{2}\sum_i (n_i-\hn)^2-J\sum_{\langle ij\rangle}(a\yd_i a\nd_j
+\mbox{H.c.})-\mu\sum_i (n_i-\hn),
\label{BHM}
\ee
where $n_i$ is the boson occupation on site $i$. 
The tunneling $J$ and interaction $U$ are known functions of the 
microscopic forces\cite{BHM}.
At integer fillings, $\hn=1,2,\ldots$,  the BHM   exhibits  quantum phase
transitions. For large tunneling (weak optical potential barriers)
$ J\hn >> U  $ the ground state is a  superfluid (Bose Condensed) with long
range phase order. Below a critical tunneling strength $J<J_c(\hn)$,  bosons
are localized in incompressible (integer  occupations) Mott phases.

In the vicinity of the Mott phase, number fluctuations are small.
An effective Hamiltonian  truncated into the subspace of lowest
local number  states $|\hn-1\rangle , |\hn\rangle ,|\hn+1\rangle$,
captures the essential correlations
around the transition. The reduced Hilbert space can be
represented by three commuting $t$-bosons $|\hn+\a\rangle=
t\yd_{\a i}|0\rangle$, $\a=1,0,-1$, which obey the holonomic
constraint $\sum_\a t\yd_{\a i}t\nd_{\a i}=1$. In this subspace,
the bosons of the BHM are represented by 
$a\yd_i=\sqrt{\hn}~t\yd_{0i}t\nd_{-1i}+\sqrt{\hn+1}~t\yd_{1i}t\nd_{0i}$.

At large $\hn$\cite{large_n}, the effective
Hamiltonian assumes a particularly simple
pseudospin-one form \bea H_{eff}&=&\frac{U}{2}\sum_i (S^z_i)^2-
J\hn\sum_{\langle ij\rangle}(S^x_i S^x_j+S^y_i
S^y_j)-\mu\sum_iS^z_i\nn\\
 S^+_i&=&
\sqrt{2}(t\yd_{0i}t\nd_{-1i}+t\yd_{1i}t\nd_{0i}) ~~~
S^z_i=t\yd_{1,i}t\nd_{1,i}-t\yd_{-1,i}t\nd_{-1,i}. \label{Heff}
\eea

The complex superfluid OP field breaks planar  spin symmetry
$\Psi   ({\bf x}_i) = \sqrt{\hn}\av{S_i^+}  $.  It is tempting to
describe the action of $H_{eff}$ using spin one coherent
states\cite{SCS}. However, the
Mott ground state     is  perturbatively connected to the O(2) rotationally
invariant state  $\prod_i |S^z_i=0\rangle $. Thus  it is difficult   to
describe this phase as a saddle point of a  spin  coherent states path
integral.

Alternatively we can use  {\em modified}
coherent states defined by
 \bea
&& \ket{\W(\theta,\eta,\phi,\chi)} =\Big[\cos (\t/2 )
t\yd_0+e^{i\h}\sin (\t/2 )\nn\\
&&~~~~~~~~~\times
\left(e^{i\f}\sin (\x/2 ) t\yd_1
+e^{-i\f}\cos (\x/2 )
t\yd_{-1}\right)\Big]\ket{0} .    \label{Omega}
\eea

{\em The mean field theory } is
similar  to previous variational approaches\cite{kotliar,caffarel}.
A  homogeneous variational wave function which   captures both phases is
$\ket{\Phi_{mf}}=\prod_i\ket{\W_i}$. 
$\theta=0$  describes  the Mott phase, while  $\theta>0$ has a
superfluid OP
$ \Psi\propto \sqrt{\hn}\sin\t$.
The variational energy per site is
\bea
e_{var}&=&\left(\frac{U}{2}+\mu\cos\x\right)\sin^2\left(\frac{\t}{2}\right)
-\frac{Jz\hn}{4}\sin^2\t \nn\\
&&\times\left(1+\hn^{-1}\sin^2(\x/2)+\sqrt{1+\hn^{-1}}\sin\x\cos2\h\right)
\eea
where $z$ is the lattice coordination number. The Mott phase boundaries
found by minimizing $e_{var}$ are given by
\be
\mu_{c}/U=-\frac{1}{8\hn u}\pm\half\sqrt{1-\frac{1}{u}\left(1+\frac{1}{2\hn}\right)
+(4\hn u)^{-2}} ~~,
\ee
where $u=U/(4J\hn z)$.
For large $\hn$ (and commensurate density) the Mott transition occurs
at $\mu=0$ and $u=1$. For $\hn=1$ it occurs at $U=5.8z J$.

The error incurred by truncation to three states per site is estimated
by comparing (\ref{Omega}) with a variational anzats
which inclues 11 occupation states. Even for $\hn>>1$ we find a wide regime
($u>0.4$) in which the probability weight of states out side
the truncated Hilbert space is less than $1\%$.
The particle number fluctuation $\d n<0.6$ in this range
and the error in $\delta n$ is less than $10\%$.

To keep the presentation simple, we focus on the limit
of large occupation numbers. This amounts to treating (\ref{Heff}),
which we believe retains the correct qualitative dynamics even for low occupations.
{\em Excitations}. Consider the canonical transformation
\bea
b_{0i}&=&\cos(\t/2)
t_0+\sin(\t/2)(t_{1i}+t_{-1i})/\sqrt{2},\nn\\
b_{\a i}&=&\sin(\t/2)
t_0-\cos(\t/2) (t_{1i}+t_{-1i})/\sqrt{2},\nn\\
b_{\varphi i}&=&(t_{1i}-t_{-1i})/\sqrt{2},
\label{bi}
\eea
with local  constraints  $\sum_m  b\yd_{m   i}b\nd_{m i}=1$.
The   mean field variational  state is simply the Fock state
$\ket{\Phi_{mf}}=\prod_i b\yd_{0i}\ket{0}$. The remaining $b\yd_\a$
and $b\yd_\varphi$ bosons create fluctuations about this state. 
The next step is to
apply the constraint  in (\ref{bi}), and eliminate $b_0$ from the
Hamiltonian\cite{comm-HP}:  \bea
b^\dagger_m  b_0&=& b^\dagger_m \sqrt{1-b\yd_\a b\nd_\a-b\yd_\varphi b\nd_\varphi}\nn\\
&\approx&
b^\dagger_m  \left( 1-\half
    b\yd_\a b\nd_\a-\half b\yd_\varphi b\nd_\varphi \ldots \right) . \eea
Truncation at quadratic order is valid provided
$\av{b\yd_\a b\nd_\a}, \av{b\yd_\varphi b\nd_\varphi}\ll 1$, 
which can  be tested  self
consistently in the regime of interest.

Now we can expand  Eq.  (\ref{Heff}) in terms of the fluctuation operators
$b_\a$ and $b_\varphi$, to obtain a harmonic hamiltonian with normal and anomalous
terms. It is diagonalized by a standard Bogoliubov transformation
to obtain
\be
H_{fluc}=   \sum_{m\vk}  \omega_{m  \vk} \beta^\dagger_{m
\vk}\beta\nd_{m, \vk} ~~~ m=1,2.
\label{Hfluc}
\ee
In the superfluid phase, there is an amplitude mode and phasons
\bea
\w_\alpha (\vk)&=&2zJ\hn\sqrt{1-u^2\gk}\approx  \sqrt{c^2\vk^2+\Delta^2},\nn\\
\w_\f (\vk)&=&Jz\hn(1+u)\sqrt{1-\gk}, \approx c |\vk|,
\label{dispersion}
\eea
where $\gk=1/z\sum_{\bf\d}e^{i\vk\cdot{\bf \d}}$, $c=J\hn\sqrt{z} (1+u)$,
$\Delta=2Jz\hn \sqrt{1-u^2}$.
The massive amplitude mode $\w_\alpha $  softens at the Mott transition and
becomes degenerate with $\w_\f$. In the Mott phase, the two modes 
\be
\w_{p}=\w_{h}=\frac{U}{2}\sqrt{1-u^{-1}\gk},
\label{wph}
\ee
are gapped, representing particle and hole excitations. 
The local density of fluctuations are found to
be relatively small $\av{b\yd_{m  i} b\nd_{m  i}}<0.08$ even at the critical
point. This  measures the accuracy of  the   variational state (\ref{Omega})
at least for the short wavelength correlations. In  the critical region,
which will be estimated later, we do not expect this mean field theory
to yield correct exponents for the vanishing of the gap or the long
wavelength correlations.

Using the modified coherent states $\ket{\W}$, we can construct a
path integral for the evolution operator as follows. A  resolution
of identity is found to be of the form:
 \be
\int_0^{\pi}d\t\int_0^{\pi}d\x\int_0^{2\pi}d\f\int_{-\pi/2}^{\pi/2}d\eta
{\cal M}(\t)\ket{\W}\bra{\W}={\cal I} ,\label{ResI-0}
\ee with the
invariant measure ${\cal M}(\t)=C  \cos\t(3\cos\t-1)$. It is also
straightforward to calculate the kinetic term
 \be
\bra{\W}\frac{d}{dt}\ket{\W}=i\sin^2(\t/2)(\dot{\h}
-\cos\x\dot{\f})\equiv -i\Upsilon(t).
\label{dWdt}
 \ee
(\ref{ResI-0})  and (\ref{dWdt}) are the necessary ingredients for
the  path integral \bea {\mathcal U}(t)=\int{\mathcal D}\W{\cal
M}(\t) \exp\Big\{i\int_0^t dt'\big[\Upsilon(t')-H(t')\big]\Big\}
\eea We now focus on the commensurate case $\mu=0$. An action in
terms of    $\Psi=  \sqrt{\hn} \sin \t e^{-i\f}$ is obtained by
integrating over the massive fields $\h$ and
$\x$. Expanding to fourth order in $|\Psi|$ and
taking the continuum limit we arrive at a ``relativistic''
Gross-Pitaevskii action
\bea S=&&\frac{1}{8J z\hn^2}\int_0^t dt' \int d^d r
\Bigg\{|\dot{\Psi}|^2
-(2J\hn)^2z|\nabla\Psi|^2 \nn\\
&&-(2J\hn z)^2(u-1)|\Psi|^2-(Jz)^2\hn u |\Psi|^4\Bigg\},
\label{action}
\eea
This  derivation is valid to second   order in the dimensionless distance to
the critical point  $ |1-u|$.

\begin{figure}[h]
\begin{center}
\vspace*{13pt}
 \leavevmode \epsfxsize1.0\columnwidth
\epsfbox{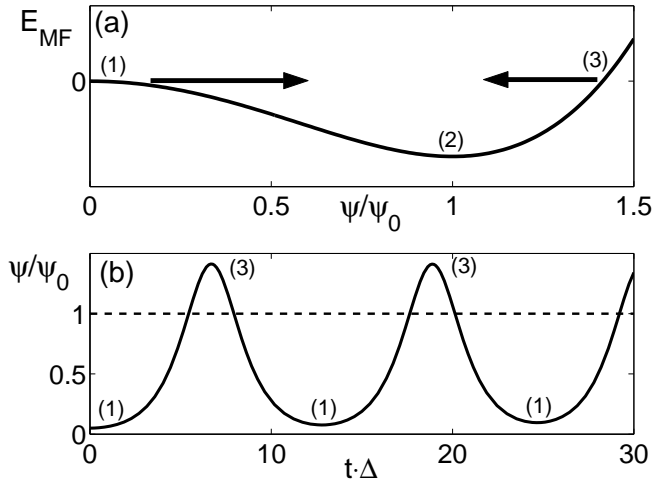}\vskip1.0pc
\caption{Oscillations of the superfluid order
parameter $\Psi$ in a system prepared in the Mott state. (a) Mean field
energy, where the equilibrium ground state is marked by point (2).
(b) Solutions of the saddle point  Eq. (\protect{\ref{eqm}}), which
exhibits macroscopic oscillations. The dashed line indicates the superfluid
OP in the ground state (2).} \label{fig:oscillate}
\end{center}
\end{figure}

{\em  Saddle point collective oscillations.}
The time evolution of the system after a sudden change of parameters
from the Mott to the superfluid state is determined by the saddle point
of the action (\ref{action}). The equation of motion for the OP, 
rescaled by its equilibrium value, $|\Psi_0|=\sqrt{2\hn (1-u)/u}$ is given by
\be
\ddot{{\bf \Psi}}=c^2\nabla^2{\bf \Psi} + \half \Delta^2{\bf \Psi} (1-|{\bf \Psi}|^2).
\label{eqm}
\ee
The constants $\D=2\sqrt{2}Jz\hn\sqrt{1-u}$ and $c=2J\hn\sqrt{z}$
are identical to the expressions below (\ref{dispersion}) to leading order in $|1-u|$.
For a uniform field configuration (\ref{eqm}) is readily integrated to give the
non linear oscillations shown in Fig. \ref{fig:oscillate}b. whose time scale is set by $\D$.
Fig. \ref{fig:oscillate}a
shows the motion of the OP on the potential landscape.

Restriction to a uniform field is not justified apriori.
In particular, topological defects may be trapped
by the so called Kibble-Zurek mechanism\cite{kibble-zurek}. However, the number of trapped
vortices can be diminished. If the initial Mott state is close to the
transition, a large correlation
length $\xi_M$ determines the distance between seed vortices.
If on the other hand we start deep in the Mott phase, we should consider
initially zero OP with small random fluctuations,
uncorrelated on the scale of a lattice constant. This amounts to a seed vortex
on almost every plaquette. However, we argue that only few will survive
the first stage of evolution.
To describe the initial growth of the OP consider the linearized version of (\ref{eqm}) around $\Psi=0$, whose eigenmodes are easily
found to be $\omega(\vk) = \sqrt{(ck)^2 - \half\Delta^2 }$.
Since it is the fastest growing, the uniform $k=0$ mode will
dominate the development of an OP. More over,
fluctuations with $k> 1/ (\xi \sqrt 2)$ do not grow at all. This implies that
defects with the same topological charge must be separated by at
least $\xi\equiv c/\D$, a large distance in the regime of interest, not
far from the transition point. The impact of the few remaining vortices on
the evolution is an interesting open issue which deserves further study.

{\em  Damping of the oscillations}.
The collective oscillations
in Fig. \ref{fig:oscillate} are
in fact a  macroscopic occupation (in a coherent state)   of the
zero wave vector amplitude mode, $\w_a({\bf k}=0)$. Since this mode is coupled
anharmonically
to the low energy phasons,   we expect a finite damping of the
oscillations due to phason pair emission.

Expanding the action in  Eq. (\ref{action}), up to the  harmonic and cubic
interactions, we obtain
\bea
S=&&\frac{1}{8Jz\hn^2}\int dt'd^dx
\Big\{\dot{\a}^2-c^2(\nabla\a)^2-\D^2\a^2\nn\\
&&+(|\Psi_0|^2+2|\Psi_0|\a)\left[\dot{\f}^2-c^2(\nabla\f)^2\right]\Big\},
\label{Scubic}
\eea
where $\a=|\Psi|-|\Psi_0|$ is the linearized amplitude mode and $c=2\sqrt{z}J\hn$.
In order to compute the damping rate of the oscillating field $\alpha$,  it is
convenient to recast the  continuum theory  in operator form,
using the amplitude and phason operators of Eq. \ref{Hfluc}
\bea
\a_{\vq=0}&=&\sqrt{\frac{2Jz\hn^2}{\D}}(\b\nd_{1,q=0}+\b\yd_{1,q=0})\nn\\
\f_k&=&\sqrt{\frac{2Jz\hn^2}{ck|\Psi_0|^2}}
(\b\nd_{2,k}+\b\yd_{2,k}).
\eea
Phason pair creation is dominated by the vertex
 \be
H_{int}=\frac{1}{\sqrt{N}}\sum_{\vk}V_{\vk}
\left(\b\nd_{1,0}\b\yd_{2,\vk}\b\yd_{2,-\vk}+\mbox{H.c.}\right).
\ee

By Fermi's golden rule, the damping rate is
\be
\Gamma=\frac{N}{(2\pi)^{d-1}}\int d^d k V(\vk)^2 \d(2ck-\D)
\ee
At wave vector $|\bar{\vk}|=\Delta/(2c)$, the vertex coupling constant is
given by \be
V_{\bar\vk}^2 =\frac{(2J\hn z)^2}{N\sqrt{2}}u(1-u)^{-\half}.
\ee
Consequently the  relative damping rate
diverges in one and two dimensions as
\be
Q_d\equiv\frac{\Gamma_d}{\D}=\frac{u}{\sqrt{2}}(1-u)^{\frac{d-3}{2}}.
\label{damping}
\ee
Oscillations could  be  observable for $\Gamma/\D<1$ which implies
$u<0.59$ for $d=1$ and $u<0.73$ for $d=2$.
In three dimensions,  the $Q_d$ is finite at the transition.

The damping ratio (\ref{damping}) can also be derived from the one
loop correction to the longitudinal susceptibility
\cite{sachdev-97}.
Higher order terms can be resummed using a large $N$
or renormalization group approach to obtain expressions
valid in the critical region\cite{sachdev-97,sachdev-99,sachdev-book}.

We comment that emitted phasons will
eventually thermalize, bringing the system to a new, finite temperature, equilibrium.
However, we are concerned with the shorter time
transient of the system, regardless of its ultimate equilibrium fate.

{\em Critical phenomena}. Time dependent  experiments
such as those proposed in Ref. \cite{orzel} could  potentially measure
quantum critical fluctuations directly. By our mean field  theory, the
collective oscillation frequency, $\w_1(k=0)$,   vanishes at the transition
as $(1-u)^{\half}$ according to Eq. (\ref{dispersion}).
We can estimate the quantum critical region
below the critical dimension $d=3$, using Ginzburg's
criterion for the $d+1$
dimensional action (\ref{action}).
For $d=2$, we have obtained $|1-u|\le 0.15$, which
overlaps with the   overdamped region  of the critical amplitude
oscillations.

In the critical  region,
the mean field gap exponents as read from  (\ref{dispersion}), and (\ref{wph})
should be modified.  To
leading order in $\e=3-d$ the gap exponent is given in Ref \cite{sachdev-book}
by: $\nu=\half+0.1\e$.

{\em Experimental parameters } for the
Boson Hubbard model extracted from  Ref.\cite{orzel},  can be translated into
$\hn\approx 50$, $2\pi \hbar/(2Jz\hn)\approx 0.7$ms. The oscillation period
is therefore larger than this timescale by a factor of $1/\sqrt{1-u^2}$.

In summary, we have described  the  dynamics
of bosons in  an optical lattice using a modified coherent states path
integral. This  affords a unified  description of both the
superfluid and Mott phases. A  system prepared in the unstable Mott state,
is expected to exhibit macroscopic
oscillations of the superfluid OP, with damping which increases
towards the transition in one and two dimensions.   It would be very
interesting to investigate superfluid   oscillations  on the Mott
side of the transition,   where there are  no low energy phason   modes
to cause damping.   Close to the transition, this would  provide
direct  measurement of the dynamical critical exponents.

After completing this paper we received a preprint by Polkovnikov etal
addressing a similar Mott to SF transition, using a dynamical
rotator representation\cite{tolia}.

{\em Acknowledgments}. We thank S. Ghosh, A. Polkovnikov and S. Sachdev for
helpful discussions. Support of the U.S.-Israel Binational Science Foundation,
and the Fund for Promotion of Research at Technion are gratefully acknowledged.

\end{document}